\documentclass[aps, pra, twocolumn, amsfonts, amsmath, amssymb, superscriptaddress, floatfix]{revtex4-2}

\usepackage[T1]{fontenc} 
\usepackage[utf8]{inputenc}  
\usepackage[english]{babel}
\usepackage{graphicx}  
\usepackage{mathbbol} 
\usepackage{kantlipsum}
\usepackage{siunitx} 
\usepackage[colorlinks,hyperindex]{hyperref}
\hypersetup{colorlinks, citecolor=blue, linkcolor=blue, urlcolor=blue}

\usepackage{bm}
\usepackage{newtxtext,newtxmath} 
\usepackage{xcolor}

\usepackage[textwidth=15mm]{todonotes} 

\usepackage{changes}
\setaddedmarkup{\textcolor{blue!80}{#1}}
\setdeletedmarkup{\textcolor{red!80}{\sout{#1}}}

\makeatletter
\def\NAT@def@citea{\def\@citea{\NAT@separator}}
\makeatother

\newcommand{\dd}{\mathrm{d}}

\newcommand{\dv}[2]{\frac{\dd #1}{\dd #2}}
\newcommand{\bra}[1]{\langle{#1}|}
\newcommand{\ket}[1]{|{#1}\rangle}
\newcommand{\braket}[1]{\langle{#1}\rangle}

\footnotetext{These authors contributed equally to this work}

\begin{document}

\title[]{Anderson Localization with Single Photons from a Quantum Emitter}


\author{Simon J. U. White$^{\S}$}
\email{simon.white@griffith.edu.au}
\affiliation{School of Mathematical and Physical Sciences, University of Technology Sydney, Ultimo, New South Wales 2007, Australia}
\affiliation{Queensland Quantum and Advanced Technologies Research Institute, Yuggera Country,   Brisbane,  Queensland,  4111  Australia}

\author{Diego N. Bernal-García$^{\S}$}
\email{dn.bernalgarcia@gmail.com}
\affiliation{Queensland Quantum and Advanced Technologies Research Institute, Yuggera Country,   Brisbane,  Queensland,  4111  Australia}

\author{Toan Trong Tran}
\affiliation{School of Electrical and Data Engineering, University of Technology Sydney, Ultimo, New South Wales 2007, Australia}

\author{Igor Aharonovich}
\affiliation{School of Mathematical and Physical Sciences, University of Technology Sydney, Ultimo, New South Wales 2007, Australia}
\affiliation{ARC Centre of Excellence for Transformative Meta-Optical Systems, University of Technology Sydney, Ultimo, New South Wales 2007, Australia}
 
\author{Alexander S. Solntsev} 
\email{alexander.solntsev@uts.edu.au}
\affiliation{School of Mathematical and Physical Sciences, University of Technology Sydney, Ultimo, New South Wales 2007, Australia}

\date{\today}

\begin{abstract}
Anderson localization of light is a fundamental emergent phenomenon in disordered systems. In arrays of coupled waveguides, it suppresses transport and causes photons to remain localized near the excitation site as coupling disorder increases. 
Here, we experimentally demonstrate Anderson localization using single photons emitted by a single-photon emitter in hexagonal boron nitride at room temperature. Despite the limited temporal coherence of the emitter, the photons undergo pronounced Anderson localization, evidenced by exponentially localized output intensity profiles in disordered waveguide lattices. 
Beyond the experimental demonstration, we develop a general theoretical framework for wave propagation in disordered tight-binding systems, showing that the configuration-averaged output intensity converges to a stationary spatial distribution at large propagation distances. In the case of off-diagonal disorder, this stationary profile is characterized by an effective localization length that exhibits a robust inverse–variance scaling with the disorder strength. 
These results establish defect-based room-temperature emitters as practical platforms for studying Anderson localization in integrated photonics and support their use in applications that exploit controlled disorder, including neuromorphic and quantum photonic architectures.
%
%
%
%
\end{abstract}

\maketitle

\section{Introduction}
\label{sec:introduction}

Wave propagation in disordered media exhibits pronounced interference effects, among which Anderson localization stands out as the suppression of transport through multiple-scattering interference. Originally introduced to describe the absence of diffusion in certain random lattices~\cite{Anderson1958}, this phenomenon has since been observed in a wide range of photonic settings. These include disordered refractive-index landscapes~\cite{Mitchell1997,Schirmacher2018,Segev2013,Capeta2011,DeRaedt1989,Gianfrate2020,Schwartz2007,christodoulides2003discretizing}, laser-written waveguide arrays~\cite{Capeta2011,Guzman2020,Martin2011,Szameit2010}, and multimode optical fibers~\cite{Karbasi2012,Mafi2019}.
Beyond optics, Anderson localization has also been observed for ultracold atomic matter waves in disordered or quasiperiodic potentials, as well as for acoustic and elastic waves and for thermal phonons in solids, where disorder can strongly suppress heat conduction~\cite{Billy2008,Roati2008,Hu2008,Luckyanova2018,Ni2021}.
These diverse realizations have motivated proposals and first demonstrations of localization-enabled devices, ranging from quantum-light transport and imaging in Anderson-localizing fibers to neuromorphic and reservoir-computing architectures that exploit multiply scattered waves in complex disordered networks~\cite{Demuth2022,MafiBallato2021,Gigan2022,Markovic2020}.
Anderson localization with few-photon light has been demonstrated using photon pairs generated by spontaneous parametric down-conversion (SPDC)~\cite{Crespi2013,DiGiuseppe2013}, revealing the impact of disorder on quantum correlations~\cite{Lahini2010, Silva2022}. However, SPDC sources are inherently probabilistic, which limits their scalability and practical use in integrated architectures without photon heralding. 
{
Deterministic solid-state quantum emitters offer a promising alternative, but a demonstration of Anderson localization using such sources has remained elusive, in part due to their limited single-photon purity, finite temporal coherence, and the challenges associated with efficient on-chip extraction.
}
%
Among solid-state platforms, single-photon emitters (SPEs) in hexagonal boron nitride (hBN) are particularly attractive, combining room-temperature operation, high brightness, narrow linewidths, and linearly polarized emission~\cite{Aharonovich2016,Tran2016,Grosso2017,Martinez2016,Dietrich2018,Nikolay2019,Jungwirth2017,Exarhos2017}, together with compatibility with cavities, fibers, and waveguides~\cite{Kim2018,Li2019,Schell2017,Vogl2017,Li2021,White2020}. Recent advances in hBN nanofabrication further enable monolithic integrated quantum photonic platforms~\cite{nonahal2023engineering}.
While localization with spatially incoherent classical light has been reported in disordered waveguide lattices~\cite{Capeta2011,Martin2011,Chriki2019}, the limited temporal coherence of solid-state emitters raises the question of whether localization survives under realistic source conditions, particularly in coupled-mode systems where temporal and spatial effects may be linked~\cite{Schiek2012}.

Here we experimentally demonstrate for the first time Anderson localization using single photons emitted by an hBN SPE at room temperature. We show that photons from an hBN SPE, despite having coherence times on the order of picoseconds~\cite{Sontheimer2017PhotodynamicsSpectroscopy}, localize robustly in laser-written disordered waveguide arrays, both at the center of an extended lattice and at its boundary. 
%
%
%
{To quantitatively interpret these observations, we develop a general theoretical framework for wave propagation in disordered tight-binding lattices. This framework shows that disorder averaging drives the output intensity toward a stationary spatial profile at large propagation distances, providing a natural basis for defining experimentally accessible localization lengths. In the case of coupling (off-diagonal) disorder, the theory predicts a simple relation between the effective localization length and the statistical properties of the disorder, enabling direct comparison between theory, simulations, and experiments.}

In this work, we demonstrate that short coherence times do not preclude Anderson localization in integrated photonic platforms and establish solid-state quantum emitters as compelling resources for on-chip quantum photonic technologies and applications that deliberately exploit controlled disorder.

\begin{figure}[!t]
    \centering
    \includegraphics[width=\linewidth]{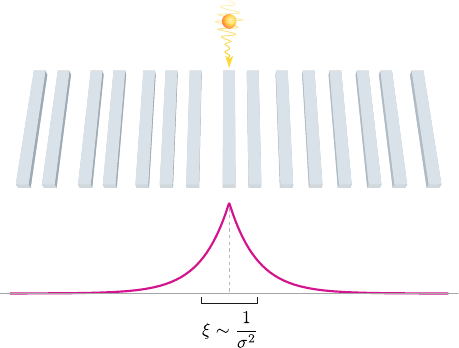}
    \caption{Schematic summary of the experiment and main theoretical result. A single photon is injected into a one-dimensional waveguide array with off-diagonal (coupling) disorder, represented by irregular inter-waveguide separations. At the output, the configuration-averaged intensity approaches a stationary exponentially localized profile characterized by an effective localization length $\xi$. For one-dimensional lattices with coupling disorder, this length obeys the inverse-variance scaling $\xi \sim 1/\sigma^2$, as given in Eq.~\eqref{eq:effective_localization}, linking disorder statistics to localization lengths extracted from end-facet measurements.}
    \label{fig:results}
\end{figure}
%

\section{Theoretical framework}
\label{sec:theoretical_framework}

Although Anderson’s original formulation concerns electronic wavefunctions in disordered solids~\cite{Anderson1958}, the same interference mechanism governs the behaviour of optical waves in coupled waveguide arrays. In this photonic setting, the slowly varying field amplitude $\psi_n$ in waveguide $n$ obeys a paraxial propagation that is mathematically equivalent to a discrete Schrödinger equation, with the propagation distance $z$ playing the role of time in a tight-binding lattice~\cite{Chen2006}. This correspondence allows light propagation to be described by
\begin{align}
    i \dv{\psi_n}{z} = \kappa_{n,n-1} \psi_{n-1} + \kappa_{n, n+1} \psi_{n+1} + \beta_n \psi_n,
\end{align}
where $\kappa_{n,n\pm1}$ are nearest-neighbor coupling constants, and $\beta_n$ are the on-site propagation constants.
It is well established that disorder in either the propagation constants or the coupling coefficients leads to Anderson localization in such lattices~\cite{Martin2011}.
%
%

{
Within this coupled-mode description, the field evolution is equivalent to the dynamics generated by a tight-binding Hamiltonian $H$, whose eigenmodes $\{\boldsymbol{\phi}_m\}$ have site amplitudes $\phi_m^{(n)}= (\boldsymbol{\phi}_m)_n$ and corresponding eigenvalues $\{E_m\}$ (see Appendix~\ref{app:hamiltonian} for the explicit form of the Hamiltonian).
The evolution generated by $H$ can then be expressed as a superposition of its eigenmodes; thus, for an excitation launched at site $n_0$, the intensity at site $n$ after a propagation distance $z$ takes the form
\begin{align}\label{eq:intensity_superposition}
    I_n(z) &= |\psi_n(z)|^2 \nonumber \\
            &= \sum_{k,\ell} \phi_k^{(n)} \phi_k^{(n_0)*} \phi_\ell^{(n)*} \phi_\ell^{(n_0)} e^{-i(E_k - E_\ell) z}.
\end{align}

%
In a disordered lattice with parameters drawn from a continuous distribution, the spectrum of $H$ is almost surely non-degenerate, and the corresponding energy gaps $E_k-E_\ell$ vanish only when $k=\ell$.
%
As a result, the cross terms with $k \neq \ell$ in Eq.~\eqref{eq:intensity_superposition} acquire rapidly oscillating phases and vanish under configuration averaging at large propagation distances.
The configuration-averaged intensity therefore converges to a stationary profile,
\begin{align}
I_{n} \equiv \lim_{z\to\infty} \langle I_{n}(z)\rangle_{\mathrm{c.a.}} = \sum_m \big\langle |\phi_m^{(n)}|^2\, |\phi_m^{(n_0)}|^2 \big\rangle_{\mathrm{c.a.}}\;,
\end{align}
which depends only on the spatial intensity profiles of the eigenmodes  of the underlying tight-binding Hamiltonian.
A detailed derivation of this result is given in Appendix~\ref{app:steady-state}.
{
Physically, this configuration-averaged profile corresponds to the dephased (diagonal-ensemble) limit of the coherent dynamics, in which interference between distinct eigenmodes is suppressed at long propagation distances due to random relative phases. 
As a result, the intensity profile becomes propagation invariant and is determined solely by the eigenmode intensities. 
%
%
This stationary behavior is a generic feature of disordered tight-binding systems with a non-degenerate spectrum, applying beyond nearest-neighbor coupling or photonic-specific assumptions.
While similar stationary intensity profiles have been observed and exploited in previous experimental and numerical studies of Anderson localization, the explicit time-dependent derivation presented here makes this limit precise and provides a unified analytical framework that enables a direct connection to effective localization length measurements.
Building on this general stationary limit, we derive below a compact and experimentally relevant relation between the measured intensity profiles and the statistics of the disorder.
}

To characterize the structure of the stationary intensity profile, we now focus on the case of purely off-diagonal disorder, which is the regime relevant for our experiment.
In the absence of on-site disorder ($\beta_n = \beta_0$), the constant term $\beta_0$ simply shifts all eigenvalues and can be absorbed into the definition of energy.
In this limit the tight-binding Hamiltonian becomes purely off-diagonal.
Further, we restrict to symmetric nearest-neighbor couplings $\kappa_{n,n+1} = \kappa_{n+1,n} \equiv \kappa_n$, so that the resulting tight-binding Hamiltonian is real and symmetric.
Borland~\cite{Borland1963} rigorously proved that all eigenstates of such Hamiltonian are exponentially localized: each eigenmode is concentrated around a random center $x_m$ and decays as
$|\phi_m^{(n)}| \propto \exp(-|n - x_m|/\xi_m)$ at sufficiently large distances, with a mode-dependent localization length $\xi_m = \xi(E_m)$~\cite{Eggarter1978}.
While the detailed site-to-site structure fluctuates between disorder realizations, the exponential decay envelope is statistically robust and captures the behavior that survives configuration
averaging. 
To obtain useful analytical expressions, we therefore adopt a phenomenological approximation in which the entire eigenmode is modeled by this exponential envelope rather than only its asymptotic tail.
Within this statistically justified description, the
dominant random variable is the eigenmode's localization center $x_m$.
Averaging over these random centers yields a stationary intensity profile that can be written as a weighted superposition of exponential decays around the launch site (see Appendix~\ref{app:effective_localization}).
For distances close to the launch site, the resulting profile is well described by a single exponential,
\begin{align}\label{eq:exponential_fit}
    I_n \simeq I_0\, e^{-2|n-n_0|/\xi},
\end{align}
which defines an \emph{effective localization length} $\xi$.
As proved in Appendix~\ref{app:effective_localization}, this $\xi$ is not the localization length of any individual eigenmode but a \emph{weighted harmonic mean} of the mode-dependent localization lengths,
\begin{align}\label{eq:harmonic_mean}
    \xi \approx
    \left(
        \frac{\sum_m w_m\,\xi_m^{-1}}
             {\sum_m w_m}
    \right)^{-1},
\end{align}
with weights $w_m = \tanh^2(1/\xi_m)\,\xi_m$ that suppress both very
weakly and very strongly localized eigenmodes.
The effective localization length therefore reflects the statistically typical localization behaviour of the ensemble.
For one-dimensional tight-binding models with purely off-diagonal disorder, the localization length of an eigenmode away from the band center is known to scale as $\xi_m \sim 1/\sigma^2$~\cite{Theodorou1976, Eggarter1978, Cheraghchi2005}, where $\sigma^2$ is the variance of the random logarithmic couplings $\ln(\kappa_n)$.
Consequently, since the weights $w_m$ restrict the average to modes with typical localization lengths, the weighted harmonic mean  in Eq.~\eqref{eq:harmonic_mean} inherits the same behavior, and the effective localization length of the stationary intensity profile in Eq.~\eqref{eq:exponential_fit} satisfies
\begin{equation}\label{eq:effective_localization}
\xi \sim \frac{1}{\sigma^2},
\end{equation}
consistent with the characteristic inverse-variance scaling of one-dimensional systems with off-diagonal disorder. 
 
The inverse–variance relation of Eq.~\eqref{eq:effective_localization} is the principal theoretical result of our study, establishing how the effective localization length depends on the statistics of the disorder. This allows $\xi$ to be estimated prior to any numerical modeling or experiment.
}
{This framework provides a quantitative link between the statistics of engineered disorder and the localization length extracted from end-facet measurements, enabling direct comparison between theory, simulations, and experiments.
In principle, the same general framework can also be applied to lattices exhibiting diagonal disorder, as well as to systems with simultaneous diagonal and off-diagonal disorder.}


\section{Numerical Modeling and Simulation}
\label{sec:Numerical_Simulation}

Having established the theoretical framework, we now describe the numerical model used to implement off-diagonal disorder in the photonic lattice and to test the predicted stationary localization behavior.

%
As noted above, we maintain $\beta_n=\beta_0$ and introduce disorder solely through random variations in the inter-waveguide spacing, which modulate the coupling coefficients $\kappa_n$.
In both simulations and fabrication, each coupling is sampled independently from a uniform distribution $\kappa_n \sim U(\kappa_0-\Delta,\,\kappa_0+\Delta)$, with $\kappa_0$ representing the nominal coupling of the ordered array, while $\Delta$ defines the disorder range. 
The dimensionless ratio $\Delta/\kappa_0$ quantifies the disorder strength and controls the transition from discrete diffraction
($\Delta/\kappa_0 = 0$) to strong localization
($\Delta/\kappa_0 \gtrsim 0.7$)~\cite{Martin2011}.
Thus, we model and fabricate arrays with $\Delta/\kappa_0 = 0.8$.

%
%

\begin{figure}[!t]
    \centering
    \includegraphics[width=0.9\linewidth]{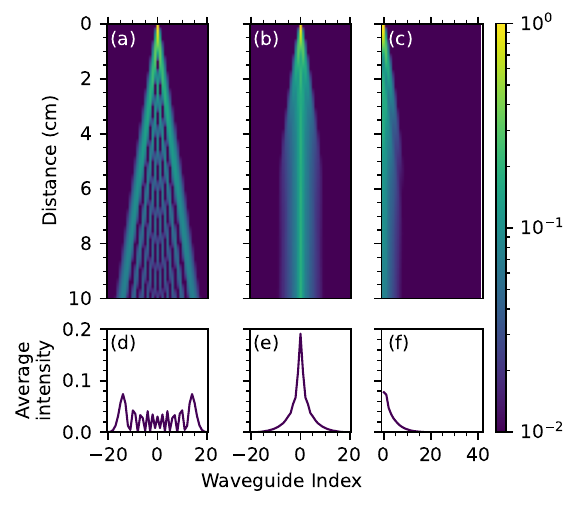}
    \caption{  Simulation of light propagation in ordered and disordered waveguide arrays. a) Discrete diffraction in an ordered array of uniformly spaced waveguides for bulk excitation. (b) Propagation in a disordered array, averaged over $10^4$ randomized disorder realizations, showing Anderson localization for bulk excitation. (c) Propagation for edge excitation in a disordered array, averaged over $10^4$ randomized disorder realizations. The external glass adjacent to the edge waveguide is modeled as an effective lossy auxiliary mode. (d)--(f) Corresponding end-facet intensity profiles after $10\,\mathrm{cm}$ of propagation for the ordered bulk-excitation case, the disordered bulk-excitation case, and the disordered edge-excitation case, respectively. Fitting the disordered end-facet profiles with Eq.~\eqref{eq:exponential_fit} yields effective localization lengths $\xi_{\mathrm{bulk}} = 5.15 \pm 0.41$ and $\xi_{\mathrm{edge}} = 7.52 \pm 0.73$, where the uncertainties denote 95\% confidence intervals.}
    \label{fig:numerical_sim}
\end{figure}

%
The impact of this controlled disorder is illustrated first through numerical simulations.
%
%
Figure~\ref{fig:numerical_sim}(a) displays the simulated evolution of the electromagnetic field in a periodic array.
An excitation launched in the central waveguide couples into its neighbors and spreads across the array, forming two outer lobes with a well-defined interference pattern in between.
The corresponding end-facet intensity profile is depicted in  Fig.~\ref{fig:numerical_sim}(d). This type of evolution in an ordered waveguide array is known as discrete diffraction. 
To observe Anderson localization, we introduce off-diagonal disorder by randomly varying the inter-waveguide separations as described above and simulate an ensemble of disordered lattices with independently drawn coupling coefficients. Exploiting the statistical shift invariance of a random lattice, we increment the launch site $n_0$ by one across realizations, enabling efficient convergence of the ensemble-averaged intensity profile.
As shown in Fig.~\ref{fig:numerical_sim}(b), the ensemble-averaged intensity evolution exhibits a clear crossover from initial discrete diffraction to pronounced localization at longer propagation distances. 
This is consistent with the stationary configuration-averaged profile discussed in Sec.~\ref{sec:theoretical_framework}.
The corresponding end-facet intensity profile [Fig.~\ref{fig:numerical_sim}(e)] displays the expected exponential decay. We also simulate excitation at the edge waveguide, where, as for bulk excitation, the wave packet becomes arrested and remains localized near the injection site, as shown in Fig.~\ref{fig:numerical_sim}(c).
%
%

The spatial intensity profiles are analyzed using the exponential form in Eq.~\eqref{eq:exponential_fit}, which has traditionally been employed as a phenomenological fit in optical Anderson localization~\cite{Schwartz2007}, but here 
is additionally supported by the stationary configuration-averaged profile derived in Sec.~\ref{sec:theoretical_framework}.

For bulk excitation, fitting the end-facet intensity profile with Eq.~\eqref{eq:exponential_fit} yields an effective localization length of $\xi_{\mathrm{bulk}} = 5.15 \pm 0.41$ [Fig.~\ref{fig:numerical_sim}(e)]. For edge excitation, the fitted value increases to $\xi_{\mathrm{edge}} = 7.52 \pm 0.73$ [Fig.~\ref{fig:numerical_sim}(f)], indicating weaker confinement near the edge.
This behavior is consistent with previous observations that edge
excitation leads to reduced localization for a fixed disorder strength in photonic lattices~\cite{Szameit2010}, reflecting the reduced number of available scattering pathways near the edge.
The propagation dynamics were computed by exact diagonalization of the tight-binding Hamiltonian and independently verified using a Crank--Nicolson scheme.
%
%

%
For the disorder distribution used here, the inverse variance of the logarithmic couplings is $1/\sigma^2 = 3.11$, which sets the expected order of magnitude of the effective localization length according to Eq.~\eqref{eq:effective_localization}. As a numerical consistency check, we also extract energy-resolved localization lengths using two standard diagnostics, namely the inverse participation ratio (IPR) of the exact eigenmodes and the transfer-matrix method (TMM), and, in each case, apply the weighted harmonic-mean prescription of Eq.~\eqref{eq:harmonic_mean} to obtain effective localization lengths $\xi_{\mathrm{IPR}} = 2.66$ and $\xi_{\mathrm{TMM}} = 2.52$, respectively. These values are consistent with the scale set by $1/\sigma^2$ and provide additional support for the inverse-variance scaling in Eq.~\eqref{eq:effective_localization} as a practical predictor of the localization length from the disorder statistics.
%

Taken together, these results support the use of Eq.~\eqref{eq:exponential_fit} as an effective description of the stationary configuration-averaged intensity profile and provide the basis for comparison with the experimental single-photon measurements presented in the next section.
%

\section{Experimental implementation}
\label{sec:experimental_implementation}
\begin{figure*}
    \centering
    \includegraphics[width=0.85\linewidth]{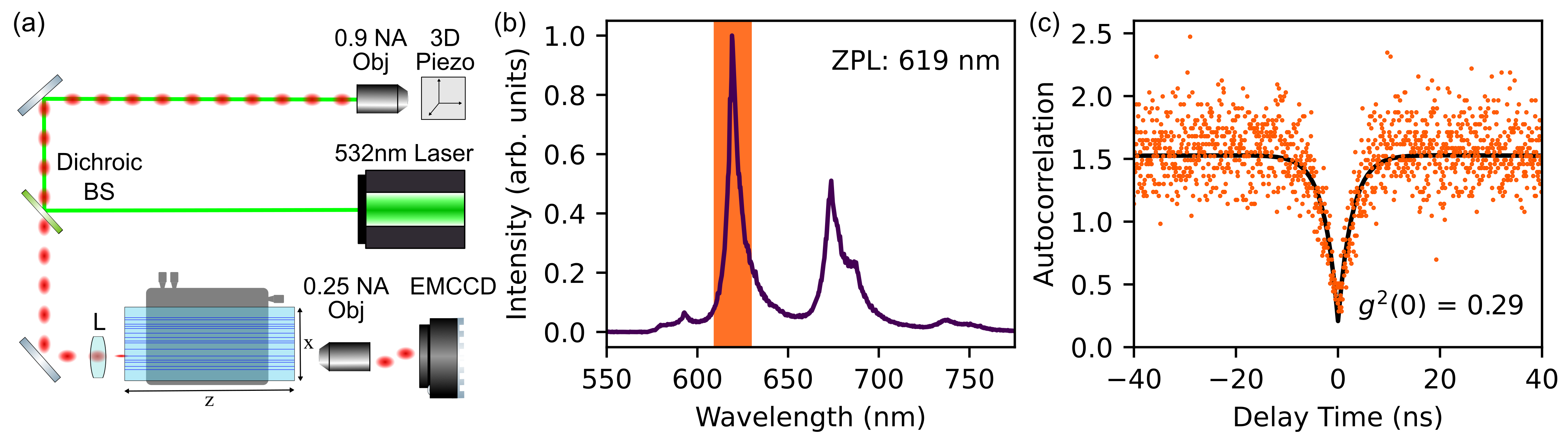}
    \caption{Single-photon source measurement. a) Experimental implementation. A hBN single photon emitter is excited with a $100\,  \mu\mathrm{W}$ $532\, \mathrm{nm}$ laser. Photons are collected with an objective and directed through an aspheric lens (L) into a single waveguide. The light propagates along the z-axis, and the end facet is imaged using an electron-multiplying CCD, capable of detecting single photons. b) hBN photoluminescence spectra with bandpass region highlighted in orange. c) Second-order correlation measurement displaying normalized $g^2(0)$ below $0.5$, without jitter correction, signifying the emission of a single defect.}
    \label{fig:experiment}
\end{figure*}
To experimentally realize the arrays of periodic and disordered waveguide arrays an integrated photonic circuit was fabricated using an established direct femtosecond laser writing technique~\cite{szameit2010discrete}. Two waveguide arrays were written inside a fused silica chip (Corning 7980), designed to observe discrete diffraction and exponential localization, respectively. In the first array, $101$ identical waveguides were homogeneously separated by $21.9\, \mu \mathrm{m}$ to realize coupling coefficients $\kappa_0 = 0.08\, \mathrm{mm}^{-1}$. The second array was designed with disorder in the coupling coefficients. Again, 101 identical waveguides were direct-laser-written, but with separations chosen such that the coupling coefficients vary within a uniform distribution of width $2\Delta$ and mean $\kappa_0$.
To observe Anderson localization with single photons, the emission from a single defect in hBN was coupled into the waveguide array, as shown schematically in Fig.~\ref{fig:experiment}(a). The defect was excited using a $100\, \mu\mathrm{W}$ $532\, \mathrm{nm}$ laser, and photons were collected using a custom confocal microscope. The photoluminescence was separated from the excitation using a $550\, \mathrm{nm}$ dichroic beam splitter, and the narrow-band zero phonon line (ZPL) at $619\, \mathrm{nm}$ was isolated with a $20\, \mathrm{nm}$ bandpass centered at the ZPL, as highlighted in Fig.~\ref{fig:experiment}(b). Fiber-based Hanbury Brown and Twiss interferometry was performed, with the second-order correlation function confirming that the defect is a single photon emitter [Fig~\ref{fig:experiment}(c)]. 
The autocorrelation data were normalized to a long delay time, and fit with the model for a three-level system 
\begin{align}
    g^{(2)}(t) = 1 - (1 + a_1)\, e^{-t/\tau_1} + a_2\, e^{-t/\tau_2}.
\end{align}
The three-level model was used to account for the photon bunching behavior commonly observed for quantum emitters in hBN under strong excitation.
The fit reveals an excited state lifetime, above saturation, of $\tau_1 = 2.7 \pm 0.1~\mathrm{ns}$, which is typical for hBN SPEs~\cite{Aharonovich2016}.
The autocorrelation dip at $g^{(2)}(t=0) = 0.29$
reveals that most of the emission originates from an independent single defect.
Finally, the single photons were directed into the selected input waveguide using free-space optics. The end facet of the arrays was imaged using an objective, and images were recorded using a single-photon resolving electron-multiplying charge-coupled device (EMCCD) camera.
\begin{figure}
    \centering
    \includegraphics[width=\linewidth]{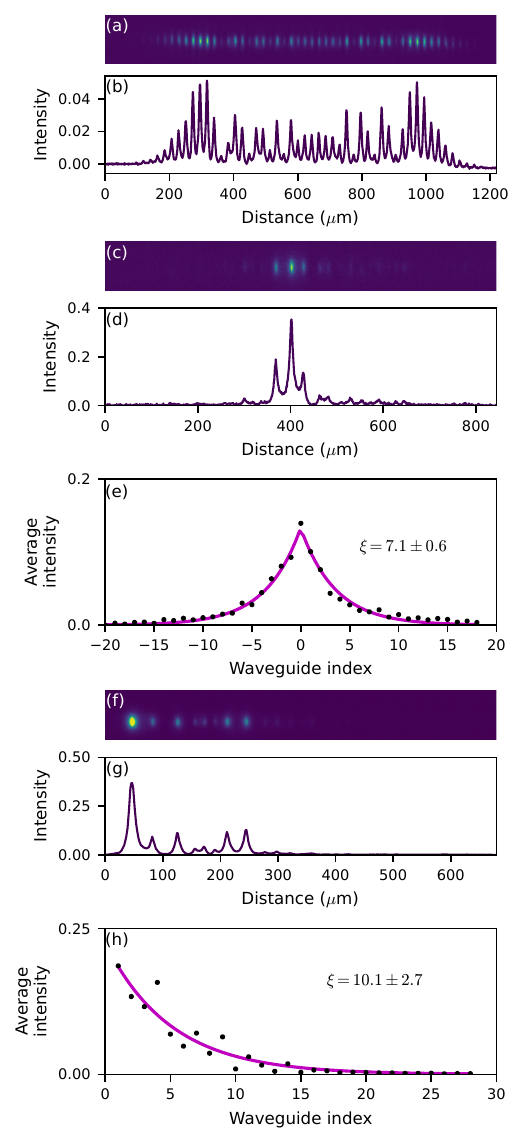}
    \vspace{-0.6cm}
    \caption{Experimental observation of single-photon propagation in periodic and disordered waveguide arrays. 
    (a,b) EMCCD image and vertically binned line scan of the chip end facet, showing diffraction through a periodic potential (equally spaced waveguides). The intensity distribution displays the conventional ballistic spread expected for discrete diffraction. (c,d) End-facet EMCCD image and corresponding line scan after propagation through a disordered array. The single-photon distribution appears localized around the injected waveguide. (e) Average intensity distribution over $30$ disorder realizations, with a localization width of $\xi = 7.1\pm0.6$. 
    (f,g) End-facet EMCCD image and line scan for single-photon injection at the boundary of a disordered lattice. For a single realization, the wave function localizes near the boundary. (h) Average intensity profile over four disorder realizations, revealing a weaker localization of $\xi=10.1\pm2.7$.
    }
    \label{fig:experimental_localization}
\end{figure}

Figure~\ref{fig:experimental_localization}(a) shows the  measured end-facet image for hBN single photons injected into the central waveguide of an array of homogeneously spaced waveguides. A vertically binned line scan, Fig.~\ref{fig:experimental_localization}(b), reveals the single-photon intensity profile diffracted through the lattice, resulting in two outer lobes and an interference pattern in between [Fig.~\ref{fig:experimental_localization}(b)], as expected from the modeling. Next, we observed the localization of the single photons through the disordered lattice. To observe the effect for a single realization of disorder, single photons were coupled into the middle of the lattice, beginning at waveguide 30 of 101. This ensured the propagation of the wavefunction through the waveguide array without reflection from the edges. As seen in Figs.~\ref{fig:experimental_localization}(c, d), the wave function localizes near the excitation site [indicated with waveguide index 0 in Fig.~\ref{fig:experimental_localization}(e)]. To observe the exponential localization, we averaged over multiple realizations of the disorder by shifting the excitation site to the neighboring waveguide and recording the intensity distribution at the end facet. This was repeated for $30$ realizations, and the peak intensity in each waveguide is averaged over all realizations, as seen in Fig.~\ref{fig:experimental_localization}(e). 
The averaged end-facet intensity profile was then fit with Eq.~\eqref{eq:exponential_fit} to extract the effective localization length.
{As described above, disorder averaging eliminates phase-sensitive interference terms and yields a stationary intensity profile determined solely by the spatial structure of the underlying eigenmodes.}

In accordance with the theoretical framework developed above, the measured end-facet intensity profile exhibits an exponential decay away from the injection site, providing direct experimental access to the effective localization length $\xi$.
Fitting the disorder-averaged bulk profile with Eq.~\eqref{eq:exponential_fit} yields $\xi = 7.11 \pm 0.55$ for the experiment (30 realizations), compared with $\xi = 5.15 \pm 0.41$ for the simulations ($10^4$ realizations).
Within the theoretical framework developed above, this experimentally extracted $\xi$ corresponds to a weighted harmonic mean of the localization lengths of the contributing eigenmodes, rather than to the localization length of any single mode.
For completeness, we also explored the behavior of single photons injected into the edge waveguide of a disordered lattice. For a single realization, the behavior is qualitatively similar to that observed for bulk excitation: disorder arrests photon diffusion through the array, and the final intensity profile remains localized at or near the injection site, as seen in Figs.~\ref{fig:experimental_localization}(f, g). 
Repeating the edge excitation for a total of four realizations and averaging the resulting end-facet intensity profiles yields Anderson localization with maximum intensity at the edge site and exponentially decaying tails for neighboring waveguides [Fig.~\ref{fig:experimental_localization}(h)]. 
Fitting the experimental averaged edge profile gives $\xi = 10.08 \pm 2.66$, compared with $\xi = 7.52 \pm 0.73$ from the edge simulation above. Owing to the smaller number of edge realizations, the corresponding experimental uncertainty is larger. 
The larger localization length for edge excitation is consistent with the weaker confinement expected near the edge.

%

%
Because the hBN SPE spectrum is relatively broad, we also measured Anderson localization using a laser whose central wavelength was scanned over the same spectral range. We repeated the analysis for 30 realizations by exciting the central waveguide of a disordered lattice with a filtered super-continuum laser ($\text{bandwidth} < 4\, \mathrm{nm}$) centered at $607$, $612$, $617$, $622$, and $627\, \mathrm{nm}$. The localization length was extracted in each case, and the comparison with the hBN single photons ($\text{bandwidth} = 7.5\, \mathrm{nm}$) is shown in Fig. 5. We observe no significant change in localization length over the measured spectrum. This further confirms that, although Anderson localization requires disorder of the coupling coefficients, the disorder-averaged dynamics over a large propagation length are not significantly affected by the wavelength variations considered here.
%

%
\begin{figure}
    \centering
    \includegraphics[width=0.75\linewidth]{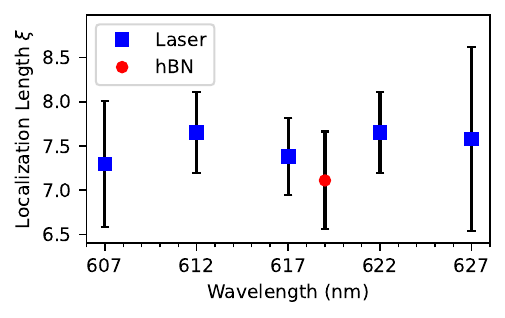}
    \caption{Localization lengths $\xi$ extracted from exponential fits to output intensity profiles for laser excitation at different wavelengths (blue squares) and for single photons emitted by an hBN defect centered at $619~\mathrm{nm}$ (red circle). The values obtained with laser excitation are consistent, within uncertainty, with those measured for hBN single photons. Error bars denote the $95\%$ confidence interval of the fitted localization width.
    }
    \label{fig:localization_width}
\end{figure}
%

%
The temporal coherence of these SPE sources typically lies in the sub-picosecond range at room temperature~\cite{White2021}, although using cryogenic cooling can approach the lifetime limit on the order of nanoseconds~\cite{fournier2023two, Fournier2023b}. To understand why this coherence is sufficient for Anderson localization, it is important to consider the characteristic distances within the waveguide array. Anderson localization depends on a photon's ability to interfere with itself as it propagates. The key property to be compared with a photon's temporal coherence is the time difference between paths a photon can take before it couples back to an initial waveguide. For the coupling disorder considered in this waveguide array, the time difference lies on average in the femtosecond range. 
This will remain true for most common photonic chip implementations without birefringence or delay-lines, including nano-photonics. This suggests more generally that single photons from typical SPE sources can undergo Anderson localization in photonic chips, provided sufficient disorder is present.
This is a crucial consideration as the potential for Anderson localization must be factored into the design of photonic integrated circuits, especially those using defect-based SPE nonclassical light sources. 
%
This observation is consistent with the theoretical result that the stationary configuration-averaged profile depends only on the eigenmode structure of the lattice Hamiltonian. 

\section{Conclusions}
\label{sec:conclusions}

{In conclusion, we have experimentally demonstrated Anderson localization of single photons emitted by a solid-state defect in hexagonal boron nitride at room temperature. 
Using laser-written waveguide lattices, we directly observed discrete diffraction in periodic arrays and exponential localization in disordered arrays, demonstrating that strong localization persists despite the limited temporal coherence of the emitter.
A comparison with spectrally filtered laser excitation further shows that the observed localization is largely insensitive to temporal coherence in the present system, confirming the suitability of room-temperature solid-state emitters for localization applications in integrated photonic lattices.
}
%
%

Beyond this experimental demonstration, we developed a general theoretical framework for wave propagation in disordered tight-binding lattices that directly connects measurable output intensity profiles to the underlying eigenmode structure.
We show that disorder averaging drives the output intensity toward a stationary, propagation-invariant profile determined solely by eigenmode intensities. 
Within this stationary regime, the effective localization length extracted from exponentially localized intensity profiles is given by a weighted harmonic mean of the localization lengths of the contributing eigenmodes [Eq.~\eqref{eq:harmonic_mean}]. 
In particular, this relation predicts an inverse–variance scaling of the effective localization length with the coupling disorder [Eq.~\eqref{eq:effective_localization}], providing a direct and experimentally relevant link between disorder statistics and localization strength.
The same general framework can also be extended to lattices with diagonal disorder or combined diagonal and off-diagonal disorder.
Our results establish defect-based quantum emitters as practical, room-temperature sources for investigating Anderson localization in integrated photonic platforms. More broadly, they provide a quantitative foundation for interpreting localization measurements in disordered photonic circuits and highlight the relevance of localization effects for applications that exploit controlled disorder, including neuromorphic photonics, inverse-designed optical systems, and reservoir-based quantum information processing.

%

\begin{acknowledgments}
We thank Friederike Klauck, Matthias Heinrich, and Alexander Szameit for their support with a waveguide array sample. 
%
Authors acknowledge the Australian Research Council (DE180100070, CE200100010, DE220100487, FT220100053) and the Office of Naval Research Global (N62909-22-1-2028) for the financial support.

\end{acknowledgments}

%
\bigskip
\appendix

\section{Intensity profile and effective localization length}
\label{app:intensity_profile}

\subsection{Hamiltonian Formulation}
\label{app:hamiltonian}

In the main text, we consider the evolution of the wavefunction in a disordered tight-binding waveguide array governed by the coupled-mode equations
\begin{equation}\label{eq:app:couple_mode}
i \dv{\psi_n}{z} = \kappa_{n,n-1} \psi_{n-1} + \kappa_{n,n+1} \psi_{n+1} + \beta_n \psi_n,
\end{equation}
where $\psi_n(z)$ denotes the field amplitude at waveguide $n$, $z$ is the propagation coordinate, $\kappa_{n,m}$ corresponds to the nearest-neighbor coupling constants, and $\beta_n$ characterizes the on-site energy of waveguide $n$.~\cite{Chen2006}

To establish the operator formalism employed in this supplemental material, we express the field amplitudes as a superposition of stationary eigenmodes,
%
\begin{equation}\label{eq:app:expansion}
\psi_n(z) = \sum_m c_m\, \phi_m^{(n)} e^{-iE_m z},
\end{equation}
where $E_m$ are the eigenvalues  (propagation constants), $\phi_m^{(n)}$ denotes the amplitude of eigenmode $m$ at site $n$, and $c_m$ is the modal weight determined solely by the initial excitation $\psi_n(0)$.
Each term in the sum satisfies Eq.~\eqref{eq:app:couple_mode} independently, leading to the time-independent tight-binding equation
\begin{equation}
\kappa_{n,n+1} \phi_{n+1}^{(m)} + \kappa_{n,n-1} \phi_{n-1}^{(m)} + \beta_n \phi_n^{(m)} = E_m \phi_n^{(m)}.
\end{equation}
%
For a system of $L$ coupled waveguides, the complete set of equations assembles into the matrix eigenvalue problem,
\begin{equation}\label{eq:app:hamiltonian_matrix}
\begin{pmatrix}
\beta_1 & \kappa_{1,2} & 0 & 0 & \cdots & 0 \\
\kappa_{2,1} & \beta_2 & \kappa_{2,3} & 0 & \cdots & 0 \\
0 & \kappa_{3,2} & \beta_3 & \kappa_{3,4} & \cdots & 0 \\
0 & 0 & \kappa_{4,3} & \beta_4 & \cdots & 0 \\
\vdots & \vdots & \vdots & \vdots & \ddots & \vdots \\
0 & 0 & 0 & 0 & \cdots & \beta_L
\end{pmatrix}
\begin{pmatrix}
\phi_1 \\ \phi_2 \\ \phi_3 \\ \phi_4 \\ \vdots \\ \phi_L
\end{pmatrix}
= E
\begin{pmatrix}
\phi_1 \\ \phi_2 \\ \phi_3 \\ \phi_4 \\ \vdots \\ \phi_L
\end{pmatrix},
\end{equation}
where we have omitted the mode index $m$ for brevity.
The tridiagonal $L\times L$ matrix in Eq.~\eqref{eq:app:hamiltonian_matrix} defines the tight-binding Hamiltonian operator $H$, whose elements encode the on-site energies and nearest-neighbor couplings.
For symmetric nearest-neighbor couplings where $\kappa_{n,n+1} = \kappa_{n+1,n} \equiv \kappa_n$, this operator can be expressed compactly as
\begin{equation}\label{eq:app:hamiltonian1}
H = \sum_{n=1}^{L-1} \kappa_n \left( |n\rangle\langle n+1| + |n+1\rangle\langle n| \right) + \sum_{n=1}^{L} \beta_n |n\rangle\langle n|,
\end{equation}
where $|n\rangle$ represents the single-excitation state localized at waveguide $n$, with $\langle n|m \rangle = \delta_{nm}$. 
The hopping term $|n\rangle\langle n+1|$ mediates amplitude transfer between adjacent waveguides, while the diagonal term $|n\rangle\langle n|$ accounts for the on-site energy.
%
%
The eigenvalue problem can then be written as $H|\phi_m\rangle = E_m|\phi_m\rangle$, whose eigenvectors take the form $|\phi_m\rangle = \sum_n \phi_m^{(n)} |n\rangle$, where $\phi_m^{(n)}= \langle n | \phi_m\rangle $ are the amplitudes introduced in Eq~\eqref{eq:app:expansion}. 
The evolution operator $e^{-iHz}$ then governs the propagation dynamics, as detailed in the following section.

\subsection{Disorder-induced steady-state intensity}
\label{app:steady-state}

We now show that, in the presence of disorder, the configuration-averaged intensity evolves toward a steady-state distribution that becomes independent of the propagation distance. This steady-state profile reflects the exponential localization of the underlying eigenmodes and defines an effective localization length characterizing the spatial decay of the mean intensity.
The field evolution in the site basis follows
\begin{equation}
\ket{\psi(z)} = e^{-iH z}\ket{\psi(0)} = \sum_n \psi_n(z)\ket{n},
\end{equation}
where $z$ is the propagation coordinate. 
To connect the site amplitudes $\psi_n(z)$ to the eigenmode expansion introduced above, we insert the spectral decomposition of the Hamiltonian into the evolution operator,
$e^{-iH z} = \sum_m e^{-iE_m z}\ket{\phi_m}\bra{\phi_m}$.
Thus, for an initial excitation localized at site $n_0$ [$\ket{\psi(0)}=\ket{n_0}$] the field amplitude at waveguide $n$ becomes
\begin{align}\label{eq:app:propagation_amplitude}
\psi_n(z) = \braket{n|\psi(z)}
&= \sum_m e^{-iE_m z}\braket{n|\phi_m}\braket{\phi_m|n_0} \nonumber \\
&= \sum_m \phi_m^{(n)}\, \phi_m^{(n_0)*}\, e^{-iE_m z}.
\end{align}
Equation~\eqref{eq:app:propagation_amplitude} expresses the field amplitude as a coherent superposition of eigenmodes, each weighted by its value at the input site and oscillating with its respective propagation constant $E_m$.
The intensity at site $n$, for an initial excitation launched at site $n_0$, is then given by
\begin{equation}\label{eq:app:intensity_sum}
I_{n,n_0}(z) = |\psi_n(z)|^2 = \sum_{k, \ell} \phi_k^{(n)} \phi_k^{(n_0)*} \phi_\ell^{(n)*} \phi_\ell^{(n_0)} e^{-i(E_k - E_\ell) z}.
\end{equation}
In a disordered lattice, the eigenvalues $E_k$ fluctuate irregularly between disorder realizations. 
As a result, the phase factors $e^{-i(E_k-E_\ell)z}$ in Eq.~\eqref{eq:app:intensity_sum} oscillate rapidly with both propagation distance and configuration, causing the cross terms ($k\neq\ell$) to average out at long distances. 
To formalize this argument, it is convenient to rewrite the configuration-averaged intensity as
\begin{equation}\label{eq:app:configuration_average}
\langle I_{n,n_0}(z)\rangle_{\mathrm{c.a.}}
   = \sum_{k,\ell}
     \big\langle 
        f_{k,\ell}^{(n,n_0)} e^{-i\Delta_{k,\ell} z}
     \big\rangle_{\mathrm{c.a.}} \equiv \sum_{k,\ell} X_{k,\ell}^{(n, n_0)}(z),
\end{equation}
where
$ f_{k,\ell}^{(n,n_0)} = \phi_k^{(n)}\phi_{k}^{(n_0)*} \phi_\ell^{(n)*}\phi_{\ell}^{(n_0)}$, $\Delta_{k,\ell}=E_k-E_\ell. $
Next, to evaluate the long-propagation limit of the configuration-averaged intensity $\langle I_{n,n_0}(z)\rangle_{\mathrm{c.a.}}$, we make use of the final-value theorem for Laplace transforms. 
Given that $X_{k,\ell}^{(n,n_0)}(z)$ is a well-behaved function of $z$, the final-value theorem for Laplace transforms gives
\begin{align}\label{eq:app:final_value}
\lim_{z\to\infty} X_{k,\ell}^{(n,n_0)}(z)
   &= \lim_{\eta\to0}
     \eta\!\int_0^\infty\! dz\, e^{-\eta z}
     X_{k,\ell}^{(n,n_0)}(z) \nonumber \\
   &= \lim_{\eta\to0} \eta\, \tilde{X}_{k,\ell}^{(n,n_0)}(\eta),
\end{align}
with the Laplace-transformed quantity defined as
\begin{equation}
\tilde{X}_{k,\ell}^{(n,n_0)}(\eta)
   = \int_0^\infty\! dz\, e^{-\eta z}
     \big\langle 
        f_{k,\ell}^{(n,n_0)} e^{-i\Delta_{k,\ell} z}
     \big\rangle_{\mathrm{c.a.}}.
\end{equation}
Equation~\eqref{eq:app:final_value} holds provided that the limit in the left-hand side exists.
It is worth noting that in the absence of disorder, the phase factors $e^{-i(E_k-E_\ell)z}$ remain perfectly coherent and, consequently, so the intensity exhibits persistent oscillations and the limit $\lim_{z\to\infty}\langle I_{n,n_0}(z)\rangle_{\mathrm{c.a.}}$ does not exist.
Exchanging the order of averaging and integration in the Laplace transform gives
\begin{equation}
\tilde{X}_{k,\ell}^{(n,n_0)}(\eta)
   = \Big\langle
      f_{k,\ell}^{(n,n_0)}
      \frac{1}{\eta+i\Delta_{k,\ell}}
     \Big\rangle_{\mathrm{c.a.}},
\end{equation}
so that, by virtue of the final-value theorem, the long-propagation limit becomes,
\begin{equation}
\lim_{z\to\infty}
   \langle I_{n,n_0}(z)\rangle_{\mathrm{c.a.}}
   = \lim_{\eta\to0}
     \sum_{k,\ell}
     \Big\langle
       f_{k,\ell}^{(n,n_0)}
       \frac{\eta}{\eta+i\Delta_{k,\ell}}
     \Big\rangle_{\mathrm{c.a.}}.
\end{equation}
In the limit $\eta\to0$, the factor $\eta/(\eta+i\Delta_{k,\ell})$ yields a Kronecker delta $\delta_{k,\ell}$, provided that the eigenmode spectrum is nondegenerate. 
This assumption is justified for independently and identically distributed (i.i.d.) disorder drawn from a continuous probability distribution, for which exact eigenvalue degeneracies occur with probability zero; thus $\Delta_{k,\ell}=0$ only when $k=\ell$.
%
%
Accordingly, all off-diagonal terms in Eq.~\eqref{eq:app:configuration_average} vanish and the configuration-averaged intensity converges to
\begin{equation}\label{eq:app:steady-state_intensity}
\bar{I}_{n,n_0} \equiv \lim_{z\to\infty}
   \langle I_{n,n_0}(z)\rangle_{\mathrm{c.a.}}
   = \sum_m
     \big\langle
       |\phi_m^{(n)}|^2
       |\phi_m^{(n_0)}|^2
     \big\rangle_{\mathrm{c.a.}},
\end{equation}
demonstrating saturation to a propagation-invariant, or \emph{steady-state}, distribution. 
This steady-state profile directly reflects the exponential localization of the underlying eigenmodes, with an envelope that decays exponentially away from the launch site, thereby defining the effective localization length $\xi$.

\subsection{Effective localization length}
\label{app:effective_localization}

We now develop a phenomenological model to determine the localization properties of the configuration-averaged intensity profile for purely off-diagonal disorder.
In the absence of on-site disorder ($\beta_n = \beta_0$), the constant term $\beta_0$ can be absorbed into the definition of energy in the eigenvalue problem, so that the Hamiltonian in Eq~\eqref{eq:app:hamiltonian1} reduces to
\begin{equation}
H = \sum_n \kappa_n \big(|n\rangle\langle n{+}1| + |n{+}1\rangle\langle n|\big),
\end{equation}
which represents a one-dimensional tight-binding model with purely off-diagonal disorder.
Borland~\cite{Borland1963} rigorously proved that all eigenstates of a Hamiltonian of this form are exponentially localized, being appreciable only within a finite region around some position $x_m$ and decaying exponentially at large distances.~\cite{Eggarter1978}
Accordingly, the spatial envelope of the $m$-th eigenmode may be  phenomenologically approximated by
\begin{equation}\label{eq:app:exponential_decay}
|\phi_m^{(n)}|^2 \approx A_m\, e^{-2|n - x_m|/\xi_m},
\end{equation}
where $\xi_m = \xi(E_m)$ denotes the localization length associated with eigenenergy $E_m$, and $A_m$ is a normalization constant depending on $\xi_m$. 
While individual eigenmodes exhibit sample-specific fluctuations, this exponential form provides a good approximation to the envelope obtained after re-centering each mode at its maximum and averaging over disorder realizations at fixed energy.
Imposing normalization, $\sum_n |\phi_m^{(n)}|^2 = 1$, yields
\begin{equation}
A_m = \tanh\!\left(\frac{1}{\xi_m}\right).
\end{equation}
%
%
%
For a fixed mode index $m$, the corresponding eigenenergies $E_m$ fluctuate only weakly around their mean value across disorder realizations, implying that the associated localization lengths $\xi_m$ also vary only slightly.
Thus, although the detailed site-to-site structure of each eigenmode differs from one realization to another, their decay envelope is statistically stable, and the main difference among realizations is the random position of the localization centers $x_m$.  
The approximation in Eq.~\eqref{eq:app:exponential_decay} should therefore be understood in a statistical sense: in the configuration average, the fine-scale fluctuations of individual eigenstates play a negligible role, and the exponential envelope provides a statistically robust description of the spatial decay.
This makes the exponential approximation a reliable basis for estimating the effective localization length that characterizes the decay of the configuration-averaged intensity.
The configuration average in Eq.~\eqref{eq:app:steady-state_intensity} requires averaging over all possible center positions $x_m$ of the exponentially localized eigenmodes with amplitudes as given in Eq.~\eqref{eq:app:exponential_decay},
\begin{align}
&\big\langle |\phi_m^{(n)}|^2 |\phi_m^{(n_0)}|^2 \big\rangle_\mathrm{c.a.} \nonumber \\
&= \frac{A_m^2}{L} \int_0^L \! \mathrm{d}x_m\,
\exp\!\left[-\frac{2}{\xi_m}\big(|n-x_m| + |n_0-x_m|\big)\right].
\end{align}
Evaluating the integral, we obtain
\begin{equation}
\label{eq:app:config_avg_single_mode}
\big\langle |\phi_m^{(n)}|^2 |\phi_m^{(n_0)}|^2 \big\rangle_\mathrm{c.a.}
   = \frac{A_m^2}{L}
     \left(\frac{\xi_m}{2} + |n-n_0|\right)
     e^{-2|n-n_0|/\xi_m},
\end{equation}
such that the configuration-averaged contribution of a single mode depends on the distance $r = |n-n_0|$ from the launch site.
Substituting Eq.~\eqref{eq:app:config_avg_single_mode} into Eq.~\eqref{eq:app:steady-state_intensity} gives
\begin{equation}\label{eq:app:superposition}
\bar{I}(r)
   \equiv \bar{I}_{n,n_0}
   = \sum_m \bar{A}_m(r)\, e^{-2r/\xi_m},
\end{equation}
where 
\begin{equation}
\bar{A}_m(r)
   = \frac{A_m^2}{L}
     \left(\frac{\xi_m}{2} + r\right)
   = \frac{\tanh^2(\xi_m^{-1})}{L}
     \left(\frac{\xi_m}{2} + r\right).
\end{equation}
For small distances $r$ relevant to the fits in the main text, the exponential factors $e^{-2r/\xi_m}$ in Eq.~\eqref{eq:app:superposition} dominate the $r$-dependence, compared with the linear variation of the prefactors $\bar A_m(r)$. 
It is therefore safe to approximate $\bar A_m(r)\simeq \bar A_m(0)$, when characterizing the local decay of $\bar I(r)$.
Under this approximation, the configuration-averaged intensity becomes,
\begin{equation}\label{eq:app:superposition2}
\bar{I}(r) 
   \simeq \sum_m \bar{A}_m(0)\, e^{-2r/\xi_m}.
\end{equation}
%
%
Equation~\eqref{eq:app:superposition2} shows that the configuration-averaged intensity profile is a weighted superposition of exponential decays, each characterized by its mode-dependent localization length $\xi_m$ and centered at the launching site.
The weights $\bar{A}_m(0)$ strongly suppress the contribution of both strongly localized modes ($\xi_m \ll 1$) and weakly localized modes ($\xi_m \gg 1$), so that modes with statistically typical values of $\xi_m$ dominate the configuration-averaged profile.
%
%

%
To quantify this behavior, we observe that the configuration-averaged intensity for small $r$ can be well approximated by a single exponential,
\begin{equation}\label{eq:app:effective_localization}
\bar{I}(r) \approx I_0\, e^{-2r/\xi},
\end{equation}
where $\xi$ defines the \emph{effective localization length} used as a single-parameter description of the dominant modes.
We determine then $\xi$ by matching the logarithmic derivatives of Eqs.~\eqref{eq:app:superposition2} and \eqref{eq:app:effective_localization} near $r=0$:
\begin{equation}
\frac{\partial}{\partial r}\ln \bar{I}(r)
   = -\frac{2}{\xi}
   = -2\,
     \frac{\sum_m w_m\, \xi_m^{-1}}{\sum_m w_m}\, ,
\end{equation}
where $w_m = \tanh^2(\xi_m^{-1})\, \xi_m$.
This yields the effective localization length as a weighted harmonic mean of the individual eigenmode localization lengths,
\begin{equation}\label{eq:app:effective_localization_harmonic_mean}
\xi
   = \left(\frac{\sum_m w_m\, \xi_m^{-1}}{\sum_m w_m}\right)^{-1}  = H_{w}[\xi_m],
\end{equation}
with weights $w_m$ that favor modes with intermediate localization lengths, in agreement with the discussion above.
For the purely off-diagonal disorder model, the asymptotic localization length away from the band center $E=0$ is known to take the form~\cite{Theodorou1976, Eggarter1978, Cheraghchi2005}
\begin{equation}
\xi(E) = \frac{g(E)}{\sigma^2},
\end{equation}
where $\sigma^2$ is the variance of the random logarithmic couplings $\ln(\kappa_n)$, and $g(E)$ is an energy-dependent factor that remains of order unity throughout the relevant part of the spectrum. 
%
%
Modes very close to the band center exhibit anomalously large localization lengths, but these modes are strongly suppressed by the weights $w_m$ in the harmonic mean and therefore do not contribute appreciably to the effective localization length $\xi$.
Substituting $\xi_m = g_m/\sigma^2$ into Eq.~\eqref{eq:app:effective_localization_harmonic_mean} yields
\begin{equation}
\xi
   = \frac{1}{\sigma^2} \left(
       \frac{\sum_m w_m\, g_m^{-1}}
            {\sum_m w_m}
     \right)^{-1}
   = \frac{H_w[g_m]}{\sigma^2},
\end{equation}
where $H_w[g_m]$ is the weighted harmonic mean of the quantities $g_m = g(E_m)$.
%
Since the weights $w_m$ suppress both very small and very large localization lengths, only modes with moderate localization contribute; for these modes, $g_m$ remains bounded within a narrow range of values close to unity.
Consequently, the harmonic mean satisfies $H_w[g_m] \sim 1$, which leads to the scaling
\begin{equation}
\xi \sim \frac{1}{\sigma^2}.
\end{equation}
Thus, the effective localization length of the stationary intensity profile exhibits the characteristic inverse–variance scaling of the one-dimensional random-hopping model.
This result corresponds to Eq.~\eqref{eq:effective_localization} in the main text.
%

%

%
%

%
%
\section{Methods}
\subsection{Sample preparation}

\paragraph{Quantum emitters in hexagonal boron nitride.}
Two drops of solvent-exfoliated hBN solution (Boron Nitride Pristine Flakes in Solution, Graphene Supermarket) were drop-cast onto a marked silicon substrate. The substrate was then annealed at $850^\circ\mathrm{C}$ under a 1-Torr argon atmosphere in a tube furnace to activate the emitters. After annealing, the sample was allowed to cool naturally to room temperature.

\paragraph{Photonic waveguide chip.}
The photonic chip was fabricated in fused silica (Corning 7980) using femtosecond direct laser writing. A commercial laser system (Coherent Mira/RegA) delivered $800~\mathrm{nm}$, $150~\mathrm{fs}$ pulses at a repetition rate of $100~\mathrm{kHz}$, which were focused $200~\mu\mathrm{m}$ below the sample surface using a $20\times$ microscope objective (numeric aperture $\mathrm{NA}=0.36$). Waveguides with a refractive-index contrast of approximately $5\times10^{-4}$ were inscribed by translating the sample relative to the laser focus using a precision positioning system (Aerotech Inc.). The resulting waveguides typically exhibit propagation losses below $0.3~\mathrm{dB/cm}$ at $815~\mathrm{nm}$ and support an elliptically shaped mode profile with minor and major-axis diameters of $\approx9$ and $\approx13~\mu\mathrm{m}$, respectively.

\paragraph{Optical characterization.}
A custom confocal microscopy and imaging setup was used to collect light from an hBN single-photon emitter and to characterize light localization in the on-chip waveguides investigated in this study. The confocal microscope employed a $532\, \mathrm{nm}$ continuous-wave green laser (Gem $532\,\mathrm{nm}$, Novanta Photonics), which was focused onto the hBN sample using a high–numerical-aperture objective (NA = 0.9, Nikon $100\times$). The position of the single-photon-emitting defect was controlled using a three-axis piezo scanner (P-611.3 NanoCube, PI) with sub-10-nm resolution. Emission from the defect was collected through the same objective, and the excitation and collection paths were separated using a $70{:}30$ (T:R) non-polarizing beamsplitter. Residual pump light was suppressed using a $532\,\mathrm{nm}$ long-pass filter (Semrock). The emitter fluorescence was subsequently filtered with a $630 \pm 10,\mathrm{nm}$ bandpass filter rotated to isolate the zero-phonon line from the phonon sideband. The emission was then routed in free space to a spectrometer, a fiber-based Hanbury Brown–Twiss setup, or into the waveguide sample, whose end facet was imaged on an electron-multiplying charge-coupled device (EMCCD).

%
\bibliography{bib}
%
%

%
%

\end{document}